\documentclass[aps,twocolumn,amsmath,amssymb, superscriptaddress]{revtex4}
\usepackage[pdftex]{graphicx}
 \usepackage{amsmath}
\usepackage[T1]{fontenc}
\usepackage{flushend}
\usepackage{amssymb}
\usepackage{amsfonts}
\usepackage{bm}
 \usepackage{amsmath} 
\usepackage{lipsum}
\usepackage{amsfonts} 
\usepackage{amssymb, mathrsfs}
\usepackage{braket}
\usepackage{graphicx} 
\usepackage{subfigure}
\usepackage{bbm}
\def\beq{\begin{equation}}
\def\eeq{\end{equation}}
\def\bsp{\begin{split}}
\def\esp{\end{split}}
\def\bea{\begin{eqnarray}}
\def\eea{\end{eqnarray}}
\def\ba{\begin{array}}
\def\ea{\end{array}}

\def\dg{\dagger}

\def\lb{\left(}
\def\rb{\right)}

\def\l.{\left.}
\def\r.{\right.}

\def\ra{\rangle}
\def\la{\langle}

\def\bo{\bold{k}}

\begin{document}

\date{\today}
\title{Noncollinear Antiferromagnetic Haldane Magnon Insulator}
\author{S. A. Owerre}
\affiliation{Perimeter Institute for Theoretical Physics, 31 Caroline St. N., Waterloo, Ontario N2L 2Y5, Canada.}
\affiliation{African Institute for Mathematical Sciences, 6 Melrose Road, Muizenberg, Cape Town 7945, South Africa.}

\begin{abstract}
In this paper we present a comprehensive study of topological magnon bands and thermal Hall effect in non-collinear antiferromagnetic systems on the honeycomb lattice  with an intrinsic  Dzyaloshinskii-Moriya interaction. We theoretically show that  the system possesses topological magnon bands with Chern number protected edge modes accompanied by a nonzero  thermal magnon Hall effect. These features result from non-collinearity of the magnetic moments due to an applied out-of-plane magnetic field.    Our results provide an experimental clue towards the realization of topological magnon transports in honeycomb antiferromagnetic compounds such as XPS$_3$ (X=Mn,Fe) and $\alpha$-Cu$_2$V$_2$O$_7$. \end{abstract}
\maketitle

\section{Introduction}

Topological magnon bands and thermal magnon Hall effect  have been realized  in the insulating quantum ferromagnets Cu(1-3, bdc)  \cite{alex6,rc}.  Previously,  thermal magnon Hall effect  was realized  experimentally in pyrochlore ferromagnets Lu$_2$V$_2$O$_7$, Ho$_2$V$_2$O$_7$, In$_2$Mn$_2$O$_7$ \cite{alex1,alex1a} following a theoretical proposal \cite{alex0, alex2}.   These materials are believed to be useful in future technological applications such as magnon spintronics.  It is generally believed that thermal magnon Hall effect  results from the nontrivial topology of magnon dispersions  \cite{alex0, alex2, zhh, alex4, alex4h,shin1,shin} encoded in the Berry curvature induced by the Dzyaloshinskii-Moriya interaction (DMI) \cite{dm, dm2}, which plays the role of spin-orbit coupling (SOC). In insulating quantum magnets the DMI is an intrinsic anisotropy and it is present due to lack of inversion symmetry of the lattice.  For honeycomb magnets, the midpoint between two magnetic ions on the next-nearest-neighbour (NNN)  bonds is not an inversion center. Thus, a DMI is allowed on the honeycomb lattice and a magnon analogue of the Haldane model \cite{fdm} can be realized in honeycomb ferromagnets \cite{sol}.  The  ferromagnetic Haldane magnon insulator also exhibits  thermal magnon Hall effect  \cite{sol1} as well as spin Nernst effect \cite{kov1,kkim}.

However, there are various antiferromagnetic materials with a honeycomb structure such as  XPS$_3$ (X=Mn,Fe)\cite{mn1,mn1a,mn2, mn3} and $\alpha$-Cu$_2$V$_2$O$_7$ \cite{Yo0,Yo1,Yo2}. The ferromagnetic  ones are rarely found. Therefore, it is very important to generalize the study of thermal magnon Hall effect to antiferromagnets (AFMs). The  antiferromagnetic systems differ from the ferromagnetic ones in various ways. The former  exhibit linear Goldstone magnon modes resulting from the spontaneous breaking of rotational symmetry. For bipartite antiferromagnets (such as the honeycomb lattice)  the magnon dispersions of the collinear N\'eel order are doubly degenerate between the $S_z=\pm S$ sectors. The degeneracy is provided by a combination of time-reversal ($\mathcal T$)  symmetry and lattice translation $T_{\bold a}$. This gives  an analogue of Kramers theorem.  The collinear antiferromagnetic  systems also have a zero net  magnetization ($\bold M$).    
 
 Motivated by the active theoretical and experimental studies of topological magnon transports,  we provide a theoretical evidence  of non-vanishing  thermal magnon Hall effect in non-collinear (canted) antiferromagnetic system on the honeycomb lattice accompanied by topological magnon bands with Chern number protected edge modes. In contrast to ferromagnets \cite{sol1,sol,kkim}, the thermal Hall response in honeycomb antiferromagnets is not induce by the DMI alone, but requires  an external magnetic field applied perpendicular to the magnet, which results in a non-collinearity of the magnetic moments  with a finite net magnetization along the field direction. This inevitably leads to topological magnon bands with Chern number protected magnon edge modes.   We  note that  the magnetic field can  also induce non-collinear N\'eel order  in frustrated honeycomb antiferromagnets such as the spin-$(3/2)$ honeycomb antiferromagnetic compound Bi$_3$Mn$_4$O$_{12}$(NO$_3$) \cite{matt, mak0, mak1, mak2, mak3, mak4}. Hence, we expect our results  to be applicable to a wide range of  frustrated antiferromagnets on the honeycomb lattice. We believe that these results will inspire  an experimental search for topological magnon bands and thermal Hall effect in honeycomb antiferromagnetic compounds.

\section{ Model }
\begin{figure}[!]
\centering
\includegraphics[width=1\linewidth]{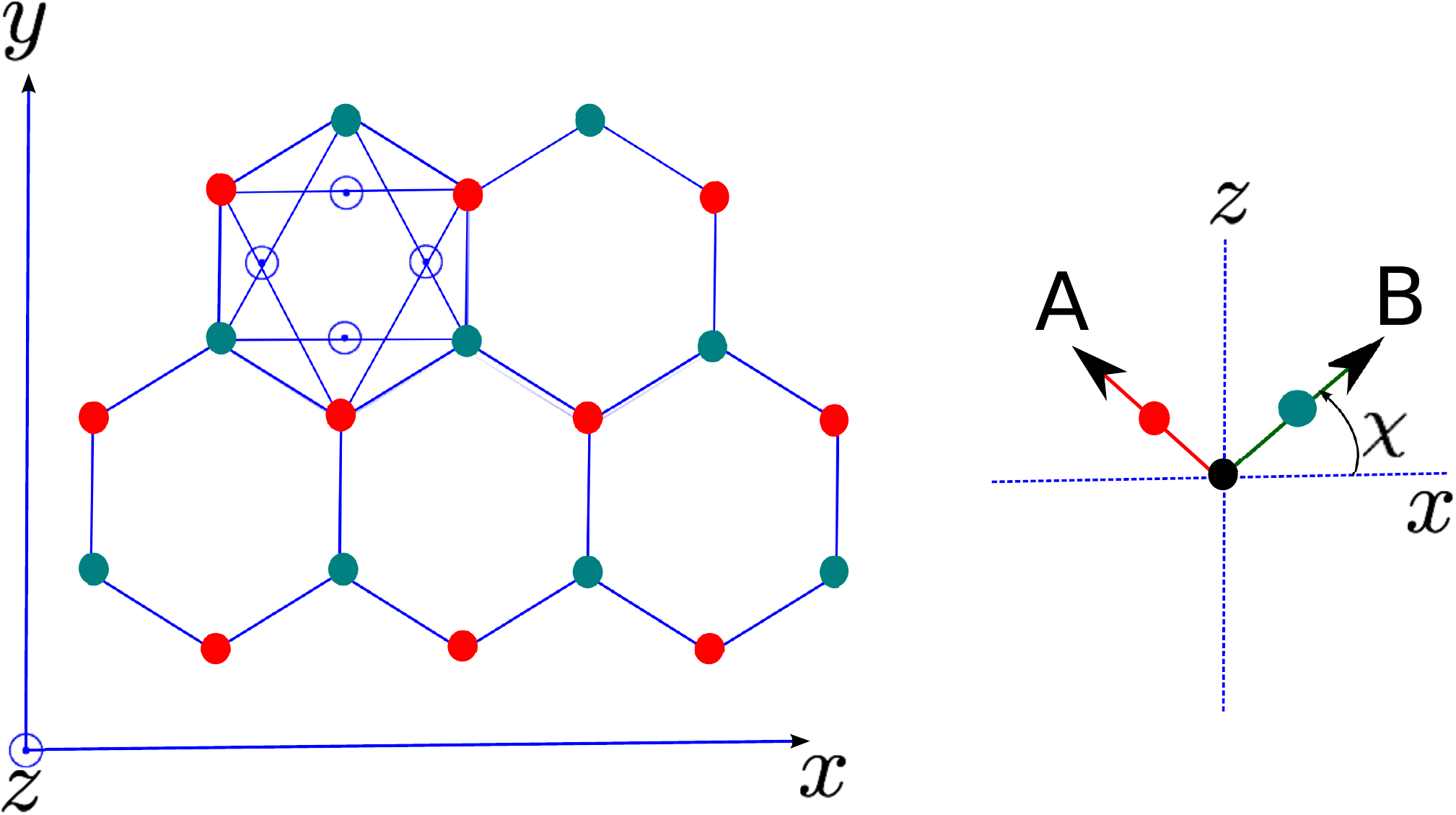}
 \caption{Color online. Schematics of the honeycomb lattice (left). The  DMI points out-of-plane at the midpoint  of the NNN  bonds as indicated by dotted circles.  A Zeeman magnetic field along the $z$-axis leads to canted N\'eel order (right), where $A$ and $B$ label the two sublattices and $\chi$ is the canting angle.}
\label{lat}
\end{figure}

We consider the antiferromagnetic quantum spin Hamiltonian on the honeycomb lattice given by 
\begin{align}
H&=J\sum_{\la i, j\ra}{\bf S}_{i}\cdot{\bf S}_{j}+\sum_{\la \la i,j\ra\ra} {\bf D}_{ij}\cdot{\bf S}_{i}\times{\bf S}_{j}-{\bold{B}}\cdot\sum_{i}\bold S_{i},
\label{model1}
\end{align}
where $J>0$ is a nearest-neighbour  antiferromagnetic interaction, ${\bf D}_{ij}=\nu_{ij}{\bf D}$ is a staggered DMI vector between sites $i$ and $j$, allowed by the NNN triangular plaquettes on the honeycomb lattice and $\nu_{ij}=\pm 1$ on sublattice $A$ and $B$ of the honeycomb lattice as shown in Fig.~\eqref{lat} with different colors similar to the Haldane model \cite{fdm}.   The last term  ${\bold{B}}=B\hat{\bold z}$ is the Zeeman interaction pointing along the $z$-axis in units of $g\mu_B$.  According to the Moriya rules \cite{dm2}, an out-of-plane DMI (${\bf D}_{ij}=\nu_{ij} D\hat{\bf z}$) is allowed due to lack of inversion center at the midpoint between the  NNN bonds as shown in Fig.~\ref{lat}. In this study, we will  consider spin-$1/2$ and set $J=1$ as the unit of energy.


\section{Magnon bands of collinear N\'eel order}
\label{coln}
Let us start with the zero magnetic field $(B=0)$ Hamiltonian  given by
\begin{align}
H&=J\sum_{\la i, j\ra}{\bf S}_{i}\cdot{\bf S}_{j}+D\sum_{\la \la i,j\ra\ra}\nu_{ij} \hat{\bf z}\cdot{\bf S}_{i}\times{\bf S}_{j}.
\label{model2}
\end{align}
At low temperatures, the ground state of \eqref{model2} is a collinear N\'eel order with the classical energy \bea E_N/NS=-\frac{3}{2}JS,\eea
where $N$ is the total number of sites. The DMI does not contribute to the classical energy  as it does not play any role in the stability of the collinear  N\'eel order at $B=0$.    At low temperatures only few magnons are thermally excited.  In this regime,  the linearized Holstein Primakoff (HP)  transformation \cite{hp} is valid. At zero magnetic field  a staggered magnetization $\bold{M}_{st}$ due to the N\'eel order can develop in any direction. However, since the DMI points out-of-plane chosen as the $z$-axis, it is natural to assume that $\bold{M}_{st}\parallel {\bf D}$ and employ the HP transformations
\begin{align} 
& S_{i,A(B)}^{z}= \pm S\mp c_{i,A(B)}^\dagger c_{i,A(B)},
\label{nel1}
\\& S_{i,A(B)}^{ y}=  i\sqrt{\frac{S}{2}}(c_{i,A(B)}^\dagger -c_{i,A(B)}),
\\& S_{i,A(B)}^{ x}=  \sqrt{\frac{S}{2}}(c_{i,A(B)}^\dagger +c_{i,A(B)}),
\end{align}
 where $c_{i,A(B)}^\dagger(c_{i,A(B)})$ are the bosonic creation (annihilation) operators and the upper and lower signs apply to   sublattice $A$ and $B$ of the honeycomb lattice.  Implementing the linear transformation above and Fourier transforming $
c_{i,A(B)}=\frac{1}{\sqrt{N}}\sum_{\bo} e^{-i\bo\cdot {\bf r}_i}c_{\bo,A(B)}$, 
 we arrive at the momentum space Hamiltonian given by 
\begin{figure}[!]
\centering
\includegraphics[width=1\linewidth]{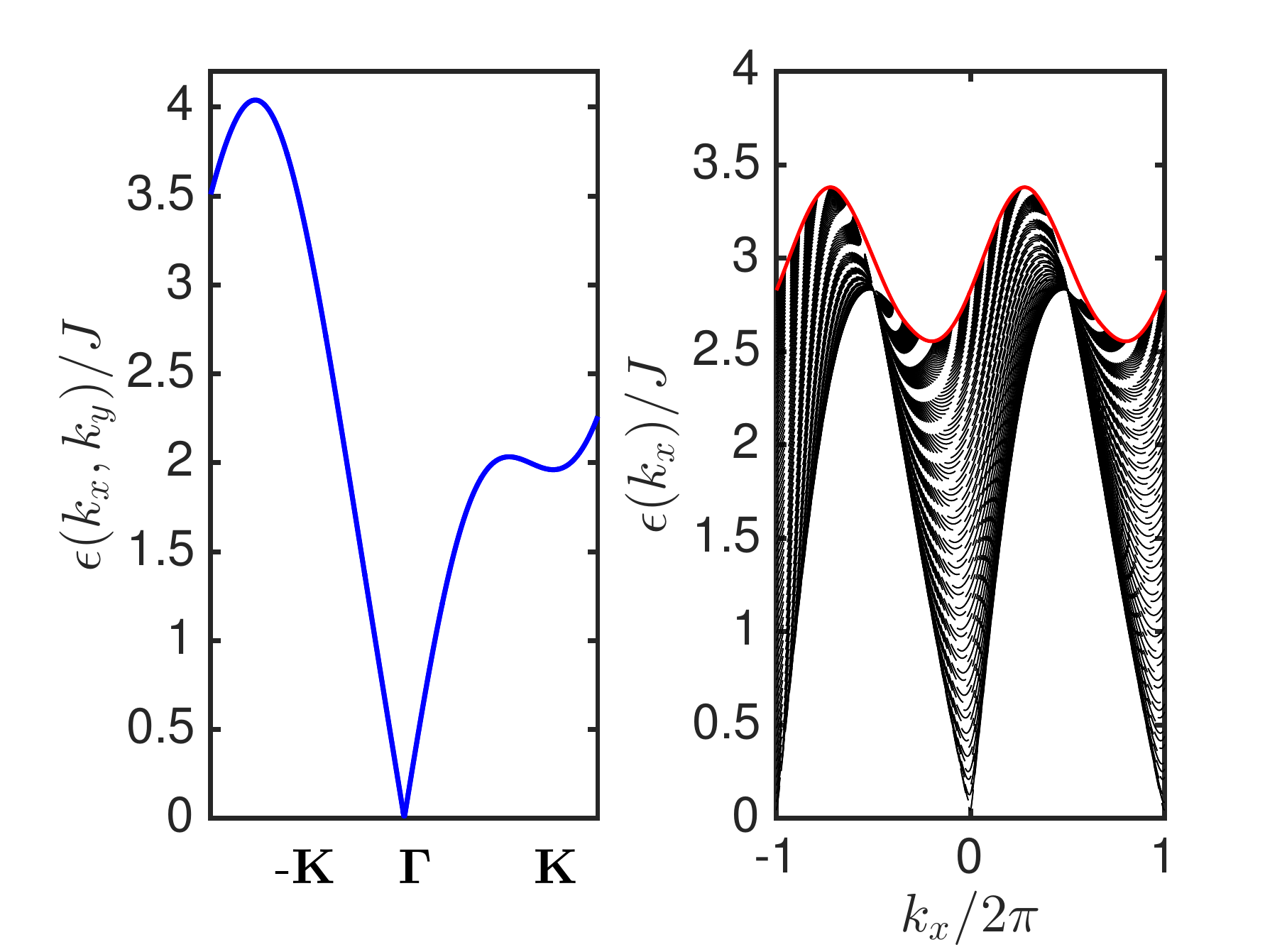}
\caption{Color online. (Left). Magnon dispersion of collinear honeycomb antiferromagnets. (Right). The corresponding magnon edge mode (red)  for a strip geometry.  $D/J=0.2$ and $B=0$.}
\label{col}
\end{figure}
 \begin{align}
\mathcal{H}_\bo=
\begin{pmatrix}
\mathcal H_{\bo I}&0\\
0&\mathcal H_{\bo II}
\end{pmatrix}.
\label{mat1}
\end{align}
The basis vector is $(c_{\bo A}^{\dg},~c_{-\bo B},~c_{-\bo A},c_{\bo B}^\dg)$,
where  $\mathcal H_{\bo II}=\mathcal H_{-\bo I}$.  
\begin{align}
\mathcal{H}_{\bo I}=
\begin{pmatrix}
I_0+m_\bo &f_\bo&\\
f^*_\bo& I_0-m_\bo 
\end{pmatrix},
\end{align}
 $I_0=3$ and 
\begin{align} 
f_\bo&=e^{ik_ya/2}\lb 2\cos(\sqrt{3}k_xa/2)+e^{-3ik_ya/2}\rb,\\
\lambda_{DM}(\bo) &= -\lambda_{DM}(-\bo)=2D\sum_i\sin\bo_i\cdot\bold{a}_i,
\end{align}
with $\bold a_1=\sqrt{3}a\hat x;~ \bold a_{2,3}=a(-\sqrt{3}\hat x, \pm 3\hat y)/2$.
The Hamiltonian \eqref{mat1} is block diagonal and each copy corresponds to ferromagnetic Haldane magnon  insulator \cite{sol,sol1}. However, due to the $\bo$ and $-\bo$ quasiparticles, we have to diagonalize $\mathcal{H}_{\bo I(II)}^\prime=\pm\sigma_z\mathcal{H}_{\bo I(II)}$, where $\pm$ applies to $I(II)$. The degenerate eigenvalues between $S_z=\pm S$ sectors  are given by 
\begin{align}
\epsilon_{\bo I}=\epsilon_{\bo II}&=\lambda_{DM}(\bo)+\sqrt{I_0^2-|f_\bo|^2},
\label{so}
\end{align}
which reflects the presence of an ``effective'' time-reversal symmetry given by $\mathcal TT_{\bold a}$. The magnon dispersion is also asymmetric since $\epsilon_{\bo I(II)}\neq \epsilon_{-\bo I(II)}$.  Figure~\eqref{col} shows the magnon dispersion and the corresponding dispersion for a strip geometry. There is a single edge mode that does not contribute to the magnon bulk dispersion, which connects states at ${\pm \bf K}=(\pm 4\pi/3\sqrt{3},0)$  on the Brillouin zone. As we will show later this system is not topological in the sense that the Chern number  vanishes.

\section{Magnon bands of noncollinear N\'eel order}
In the presence of an external  magnetic field  the N\'eel order is no longer collinear but cant in the direction of the field.  In the classical limit the spin operators  can be approximated as classical vectors,  written as
 $\bold{S}_{i}= S\bold{n}_i$, where $\bold{n}_i=\lb\cos\chi\cos\theta_i, \cos\chi\sin\theta_i,\sin\chi \rb$,
 is a unit vector and $\theta_i=\theta_{A(B)}$ labels the spin oriented angles on each sublattice with $\theta_{A(B)}=0(\pi)$, and  $\chi$ is the field-induced canting angle. The classical energy is given by
\begin{align}
E_{\text{Neel}}^{\text{cant}}/NS=-\frac{3}{2}JS\cos2\chi- B\sin\chi.
\end{align}
 Minimizing the classical energy yields the canting angle $\sin\chi= B/B_s$ with $B_s=6JS$. For the magnetic excitations above  the classical ground state, we have to redefine the geometry of the system such that the N\'eel order is on the $x$-$y$ plane at zero magnetic field and  cant slightly along the $z$-axis at finite magnetic field.  Therefore, we   have to   rotate  the coordinate axes such that the $z$-axis coincides with the local direction of the classical polarization \cite{ mov, mak1}.  The appropriate rotation on the two sublattices is given by 
\begin{align}
&S_{i, A(B)}^x=\pm S_{i, A(B)}^{\prime x}\sin\chi   \pm S_{i, A(B)}^{\prime z}\cos\chi,\label{tt}\\&
S_{i, A(B)}^y=\pm S_{i, A(B)}^{\prime y},\\&
\label{tt1}
S_{i, A(B)}^z=- S_{i, A(B)}^{\prime x}\cos\chi + S_{i, A(B)}^{\prime z}\sin\chi,\\\nonumber
\label{tt2}
\end{align}
where the primes denote the rotated coordinate and $\pm$ applies to sublattices $A$ and $B$ respectively.  The rotated Hamiltonian takes the form \begin{align}
&H_{J}^{\prime}= J\sum_{\la i,j\ra}[\cos 2\chi (S_{i}^{\prime x}S_{j}^{\prime x}-S_{i}^{\prime z}S_{j}^{\prime z})-S_{i}^{\prime y}S_{j}^{\prime y}],\\&
H_{DM}^{\prime}= D\sin\chi\sum_{\la\la i, j\ra\ra}\nu_{ij}\hat{\bold z}\cdot {\bf S}_{i}^{\prime}\times {\bf S}_{j}^{\prime}\label{dm1},\\&
H_{Z}^{\prime}= -B \sin\chi\sum_i S_{i}^{\prime z}.
\end{align}
We have dropped terms that would give magnon-magnon interaction as they will not contribute to the low-temperature physics.  In the basis vector $(\psi_\bo^\dg,\psi_{-\bo})$, where $\psi_\bo^\dg=(c_{\bo,A}^\dg,c_{\bo,B}^\dg)$, the momentum space Hamiltonian is given by
\begin{align}
H=\frac{1}{2}S\sum_{\bo} (\psi_\bo^\dg,\psi_{-\bo})\mathcal{H}_{\bo}{\psi_\bo\choose \psi_{-\bo}^\dg} +\mathcal{E}_0,
\end{align}
where $\mathcal{E}_0$ is a constant and $v_\chi=\sin^2\chi$. The momentum space Hamiltonian is 
\begin{widetext}
\begin{align}
\mathcal{H}_{\bo}=
\begin{pmatrix}
I_{0}-\lambda_{DM}^{\prime}(\bo) &-v_\chi f_{\bo}^* &0&(1-v_\chi)f_{\bo}^*\\
-v_\chi f_{\bo}&I_{0} +\lambda_{DM}^{\prime}(\bo) &(1-v_\chi)f_{\bo}&0\\
0&(1-v_\chi)f_{\bo}^*&I_{0} +\lambda_{DM}^{\prime}(\bo)&-v_\chi f_{\bo}^* \\
(1-v_\chi)f_{\bo}&0&-v_\chi f_{\bo}&I_{0} -\lambda_{DM}^{\prime}(\bo)
\end{pmatrix}.
\label{outp}
\end{align}
\end{widetext}
  The rotated mass is $\lambda_{DM}^{\prime}(\bo)=  \lambda_{DM}(\bo)\sqrt{|v_\chi|}$. The Hamiltonian is diagonalized by the generalized Bogoliubov  transformation
\begin{figure}[!]
\centering
\includegraphics[width=1\linewidth]{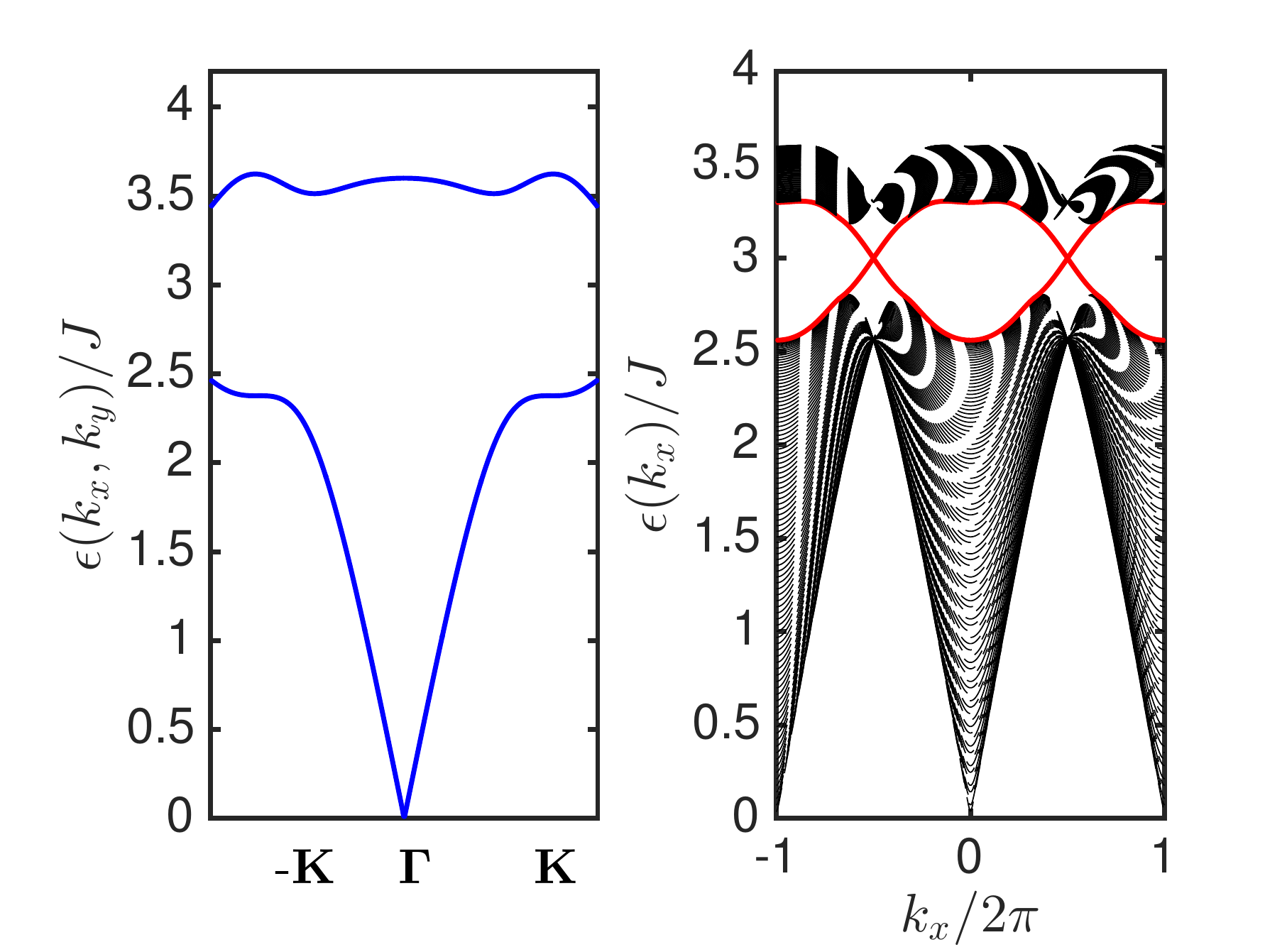}
\caption{Color online. (Left) Magnon dispersions of the canted honeycomb antiferromagnets. (Right) The corresponding  edge modes (red). $D/J=0.2$ and $B=0.3B_s$.}
\label{noncol}
\end{figure}
\begin{align}
 {\psi_\bo\choose \psi_{-\bo}^\dg}= \mathcal{U}_\bo {\Psi_\bo\choose \Psi_{-\bo}^\dg}
\end{align}
where $\Psi_\bo^\dg=(\gamma_{\bo,A}^\dg,\gamma_{\bo,B}^\dg)$ and $\gamma_{\bo,A(B)}^\dg$  are the Bogoliubov quasiparticles.  $\mathcal{E}_\bo= \mathcal{U}_\bo^\dg\mathcal H_\bo\mathcal{U}_\bo=\text{diag}(\epsilon_{\bo}, \epsilon_{-\bo}).$  The explicit form of $\mathcal{U}_\bo$ is given by
 \begin{align}
& \mathcal{U}_\bo= \begin{pmatrix}
  u_\bo& -v_\bo^* \\
-v_\bo&u_\bo^*\\  
 \end{pmatrix},
 \label{para}
\end{align} 
where $u_\bo,~v_\bo$ are  $N\times N$ matrices that satisfy \bea|u_\bo|^2-|v_\bo|^2={\bf I}_{N\times N},\eea
where ${\bf I}_{N\times N}$ is  $N\times N$ identity matrix.
The matrix $\mathcal{U}_\bo$ is paraunitary, it satisfies the relation $\mathcal{U}_\bo^\dg \eta \mathcal{U}_\bo= \eta$,  with $\eta=\text{diag}(\bold I_{N\times N},-\bold I_{N\times N})$. Using the fact that $\mathcal U_\bo^\dg=\eta \mathcal U_\bo^{-1}\eta$ and $\mathcal U_\bo^{-1}\mathcal U_\bo={\bf I}_{N\times N}$, we have 
\bea 
\eta\mathcal{H}_\bo \mathcal{U}_\bo= \mathcal{U}_\bo\eta\mathcal{E}_\bo.
\label{eig}
\eea
  Hence, the problem of finding the matrix $\mathcal{U}_\bo$  is equivalent to diagonalizing the Hamiltonian $\mathcal H_\bo^{\prime}=\eta\mathcal H_\bo$ whose eigenvalues are $\eta\mathcal{E}_\bo$ and the columns of $\mathcal{U}_\bo$ are the corresponding eigenvectors. The magnon bands are given by
   \begin{align}
\epsilon_{\bo\pm }=\sqrt{I_0^2+\lambda_{DM}^{\prime 2}(\bo)-(1-2v_\chi)|f_\bo|^2\pm 2 I_0 g_\bo}
\label{hl}
\end{align}
where $g_\bo=\sqrt{\lambda_{DM}^{\prime 2}(\bo)+|v_\chi f_\bo|^2}$.
Evidently, a finite magnetic field breaks the degeneracy between the $S_z=\pm S$ N\'eel states and a uniform net magnetization is also induced along the field direction, i.e., $M_z=\sum_{i;\alpha=A,B}\la S_{i,\alpha}^z\ra\neq 0$ from Eq.~\ref{tt1}. The magnon dispersions are shown in Fig.~\eqref{noncol}. The right column of Fig.~\eqref{noncol} shows the dispersion for a strip geometry, which clearly shows  gapless magnon edge modes. This reflects a non-collinear antiferromagnetic topological Haldane magnon insulator with nonzero Chern numbers (see the next sections). Near $\bo= {\bf \Gamma}$ the lowest energy dispersion is
    \begin{align}
\epsilon_{\bo\to {\bf \Gamma}, -}\approx 3\sqrt{\frac{1}{2}(1-v_\chi) \bo^2 + \frac{1}{16}v_\chi\bo^4}.
\label{hl1}
\end{align}
In the non-collinear (canted) antiferromagnetic phase $v_\chi$ is small, Eq.~\ref{hl1} becomes a linear Goldstone mode. Near the collinear ferromagnetic phase $v_\chi$ is close to unity and Eq.~\ref{hl1} becomes a quadratic Goldstone mode. These differences will show up later in the thermal Hall conductivity at low temperatures. 
  Near $\bo = {\bf K}$, $| f_\bo|\to 0$, we get
    \begin{align}
\epsilon_{\bo\to{\bf K},\pm}\to |I_0 \pm \lambda_{DM}^{\prime}(\bo\to {\bf K})|.
\label{hl2}
\end{align}
The Dirac magnon energy gap is given by $\Delta_{\bo\to {\bf K}}=2|\lambda_{DM}^{\prime}(\bo\to {\bf K})|=6D\sqrt{3|v_\chi|}$.

 The components of the magnetic moments  perpendicular to the field in Fig.~\ref{lat} form a N\'eel order  with vanishing magnetization $M_x=\sum_{i;\alpha=A,B}\la S_{i,\alpha}^x\ra= 0$.  The resolution of the DMI along this direction gives $
  H_{DM}^\prime= D\cos\chi\sum_{\la\la i, j\ra\ra}\hat{\bold z}\cdot {\bf S}_{i}^\prime\times {\bf S}_{j}^\prime$, which also points perpendicular to the bonds as expected. The corresponding momentum space Hamiltonian and magnon bands are given by
\begin{widetext}
\begin{align}
\mathcal{H}_{\bo}=
\begin{pmatrix}
I_{0}+\lambda_{DM}^{\prime\prime}(\bo) &-v_\chi f_{\bo}^* &0&(1-v_\chi)f_{\bo}^*\\
-v_\chi f_{\bo}&I_{0} +\lambda_{DM}^{\prime\prime}(\bo) &(1-v_\chi)f_{\bo}&0\\
0&(1-v_\chi)f_{\bo}^*&I_{0} -\lambda_{DM}^{\prime\prime}(\bo)&-v_\chi f_{\bo}^* \\
(1-v_\chi)f_{\bo}&0&-v_\chi f_{\bo}&I_{0} -\lambda_{DM}^{\prime\prime}(\bo)
\end{pmatrix},
\label{inp}
\end{align}
\end{widetext}
\begin{align}
\epsilon_{\bo,\pm}&=\lambda_{DM}^{\prime\prime}(\bo)+\sqrt{I_0^2-(1-2v_\chi)|f_\bo|^2 \pm 2 I_0 v_\chi|f_\bo|},
\label{inpe}
\end{align}
where  $\lambda_{DM}^{\prime\prime}(\bo)=  \lambda_{DM}(\bo)\sqrt{1-v_\chi}$. It is evident that Eq.~\ref{inpe} recovers Eq.~\ref{so} at  zero magnetic field $v_\chi=0$.  The lowest energy dispersion of \eqref{inpe} also has a linear Goldstone mode  near $\bo={\bf \Gamma}$ for small $v_\chi$. However, near $\bo = {\bf K}$ we get
    \begin{align}
\epsilon_{\bo\to{\bf K},\pm}=I_0+ \lambda_{DM}^{\prime\prime}(\bo\to {\bf K}),
\label{hl2}
\end{align}
where $\lambda_{DM}^{\prime\prime}(\bo\to {\bf K})=3D\sqrt{3(1-v_\chi)}$. The Dirac magnon gap  between the top and bottom bands vanishes $\Delta_{\bo\to {\bf K}}=0$.  The absence of gap Dirac dispersion at $\bo = {\bf K}$ shows that the components of the magnetic moments  perpendicular to the field do not form an antiferromagnetic topological magnon insulator.

 \section{Berry curvature and Hall response}
The basic idea of a nontrivial topological system is that  the Berry curvature is finite and the corresponding Chern number takes integer values. This idea also applies to insulating quantum magnets although the excitations are bosonic quasiparticles.  The Berry curvature is defined using the eigenstates of the system. In the present model the eigenstates are encoded in the   paraunitary operator $\mathcal U_{\bo}$ that diagonalizes the spin wave Hamiltonian.  Hence, the Berry curvature can be written as 
 \begin{align}
 \Omega_{ij,\alpha}(\bo)=-2\text{Im}[\eta\mathcal (\partial_{k_i}\mathcal U_{\bo\alpha}^\dg)\eta(\partial_{k_j}\mathcal U_{\bo\alpha})]_{\alpha\alpha},
 \label{bc1}
 \end{align}
 where $\alpha$ labels the components.   Using  the relation between  $\mathcal{U}_{\bo\alpha}$ and $\eta\mathcal{H}_\bo$ as shown above, the Berry curvature can be written alternatively as
 \begin{align}
\Omega_{ij;\alpha}(\bold k)=-\sum_{\alpha\neq \alpha^\prime}\frac{2\text{Im}[ \braket{\mathcal{U}_{\bo\alpha}|v_i|\mathcal{U}_{\bo\alpha^\prime}}\braket{\mathcal{U}_{\bo\alpha^\prime}|v_j|\mathcal{U}_{\bo\alpha}}]}{\lb\epsilon_{\bo\alpha}-\epsilon_{\bo\alpha^\prime}\rb^2},
\label{chern2}
\end{align}
where   $v_{i}=\partial [\eta\mathcal{H}_\bo]/\partial k_{i}$ defines the velocity operators. The Chern number is given by
\begin{equation}
\mathcal{C}_\alpha= \frac{1}{2\pi}\int_{{BZ}} dk_idk_j~ \Omega_{ij; \alpha}(\bold k).
\label{chenn}
\end{equation} 

 In electronic system an electric field $\boldsymbol {\mathcal E}$ induces a charge current $\boldsymbol {\mathcal J}_e$, and the Berry curvature acts as an effective magnetic field giving rise to quantum anomalous Hall effect. For insulating quantum magnets the bosonic excitations are charge-neutral, which do not feel a Lorentz force. However,  a temperature gradient $-\boldsymbol { \nabla} { T}$ can induce a heat current $\boldsymbol {\mathcal J}_Q$, and the Berry curvature also acts as an effective magnetic field giving rise to a non-quantized thermal magnon Hall effect. From linear response theory, one obtains $\mathcal J^{\alpha}_Q=-\sum_{\beta}\kappa_{\alpha\beta}\nabla_{\beta} T$, where $\kappa_{\alpha\beta}$ is the thermal conductivity and $\alpha,\beta$ label the components.  The transverse component $\kappa_{xy}$ is associated with thermal Hall conductivity given explicitly in Ref.~\cite{shin1}, which has the form
\begin{align}
\kappa_{xy}=-\frac{k_B^2 T}{\hbar V}\sum_{\bo}\sum_{\pm} c_2[ f_{BE}\lb\epsilon_{\bo \pm}\rb]\Omega_{xy,\pm}(\bold k), 
\label{thm}
\end{align}
where $V$ is the volume of the system.  
\bea 
f_{BE}(\epsilon_{\bo\pm})=\lb e^{{\epsilon_{\bo\pm}}/k_BT}-1\rb^{-1},
\label{bose}
\eea is the Bose function, $k_B$ is the Boltzmann constant, $T$ is the temperature.  
\bea 
c_2(x)=(1+x)\lb \ln \frac{1+x}{x}\rb^2-(\ln x)^2-2\text{Li}_2(-x),
\eea 
and $\text{Li}_2(x)$ is a dilogarithm. 
\subsection{Collinear N\'eel order}

At zero magnetic field the Hamiltonian is block diagonal.
To obtain the eigenvectors, we solve the eigenvalue equation, $ \mathcal{H}_{\bo I(II)}^\prime \tilde{\mathcal{U}}_{\bo }=\epsilon_{\bo I(II)}\tilde{\mathcal{U}}_{\bo},$ where $\tilde{\mathcal{U}}_{\bo}$ is the eigenvector corresponding to the positive eigenvalue $\epsilon_{\bo I}$, which is given by the second column  of  $\mathcal{U}_{\bo}$, {\it i.e.,} $\tilde{\mathcal{U}}_{\bo}={-v_\bo^*\choose u_\bo^*}$.  We find
\begin{align}
u_\bo&=e^{i\phi_\bo}\cosh\lb\frac{\theta_\bo}{2}\rb,\quad
v_\bo=\sinh\lb\frac{\theta_\bo}{2}\rb,
\end{align}
where
 \begin{align}
 \cosh\theta_\bo&=\frac{I_0}{\omega_\bo};~ \sinh\theta_\bo=\frac{|f_\bo|}{\omega_\bo};~
 \tan\phi_\bo= \frac{\textrm{Im} f_\bo}{\textrm{Re} f_{\bo}},
 \end{align}
 and $\omega_\bo=\sqrt{I_0^2-|f_\bo|^2}$.  The Berry curvature for both energy branches  from Eq.~\ref{bc1} with $\eta=\sigma_z$, is given by
\begin{align}
\Omega_{ij;\alpha}(\bo)&=2\text{Im}[\partial_{k_i}u_\bo \partial_{k_j}u_\bo^*]\cdot (\sigma_z)_{\alpha\alpha},
\label{bee}
\end{align}
where  the term $\partial_{k_i}v_\bo \partial_{k_j}v_\bo^*$ is real and $i,j=\lbrace x,y\rbrace$. Differentiating Eq.~\ref{bee} yields
\begin{align}
\Omega_{ij;\alpha}(\bo)&=\frac{\sinh\theta_\bo }{2}[\partial_{k_i}\phi_\bo\partial_{k_j}\theta_\bo-\partial_{k_j}\phi_\bo\partial_{k_i}\theta_\bo ]\cdot (\sigma_z)_{\alpha\alpha}.
\end{align}
Hence, the Berry curvature is independent of the DMI and it is equivalent to that of honeycomb antiferromagnet without DMI \cite{sol2}.  Although  the Berry curvature is nonzero, the Chern number $\mathcal{C}_\alpha$ vanishes at zero magnetic field due to $\mathcal TT_{\bold a}$-symmetry. Another property of this system is that thermal spin Hall response also vanishes at zero magnetic field as we now discuss.  At zero magnetic field the diagonal magnon dispersion comprises $\epsilon_{\bo \alpha}$ and $\epsilon_{-\bo \alpha}$, therefore we have  to add the two contributions in Eq.~\eqref{thm} as they are not equal to each other. The contribution from the DMI  cancels out and the resulting  transverse thermal Hall conductivity vanishes, $\kappa_{xy}=0$. The vanishing Hall response in collinear antiferromagnets is due to  $\mathcal TT_{\bold a}$-symmetry, which  gives an analog of Kramers degeneracy, but  thermal spin Nernst response  persists \cite{kov,ran}.   This is analogous to the vanishing of quantum anomalous Hall conductivity in Kane-Mele model \cite{yu3}.  This result can also be understood by means of zero net magnetization of collinear antiferromagnets. 

\subsection{Noncollinear N\'eel order}  Recent  experiments  have observed  thermal magnon Hall effect \cite{alex1,alex1a, alex6} and topological magnons \cite{rc} in ferromagnets.  Thus, the realization of  topological magnons and  thermal magnon Hall effect in honeycomb (anti)ferromagnets will be of interest in quantum magnetism. Here, we show that thermal magnon Hall effect  does not vanish in non-collinear honeycomb antiferrmomagnets.  A nonzero magnetic field can induce a uniform net magnetization  along the field direction and a staggered magnetization  along perpendicular to the field direction. As shown above the former form a non-collinear antiferromagnetic Haldane magnon insulator characterized by a Chern number of $\mathcal C_{\pm}=\pm 1$ for the upper and lower magnon bands.  The latter, however,  is not topological because the Berry curvature \eqref{chern2} vanishes by symmetry.   Therefore, we expect that thermal magnon Hall effect is defined only in the former case. Unfortunately, explicit analytical calculations are very tedious, so we will resort to numerical integration in this section.
We first compute the  average sublattice  magnetization,  given by \cite{kle1}
\begin{align}
\la \bold{S}_i\ra =(SN-\sum_\bo \sum_{n=3}^4 |\mathcal{U}_{\bo,\ell,n}|^2)\bold{n}_{i},
\end{align}
where $\ell= 1, 2$ for the sublattice $i=A,B$ respectively.
\begin{figure}
\centering
\includegraphics[width=1\linewidth]{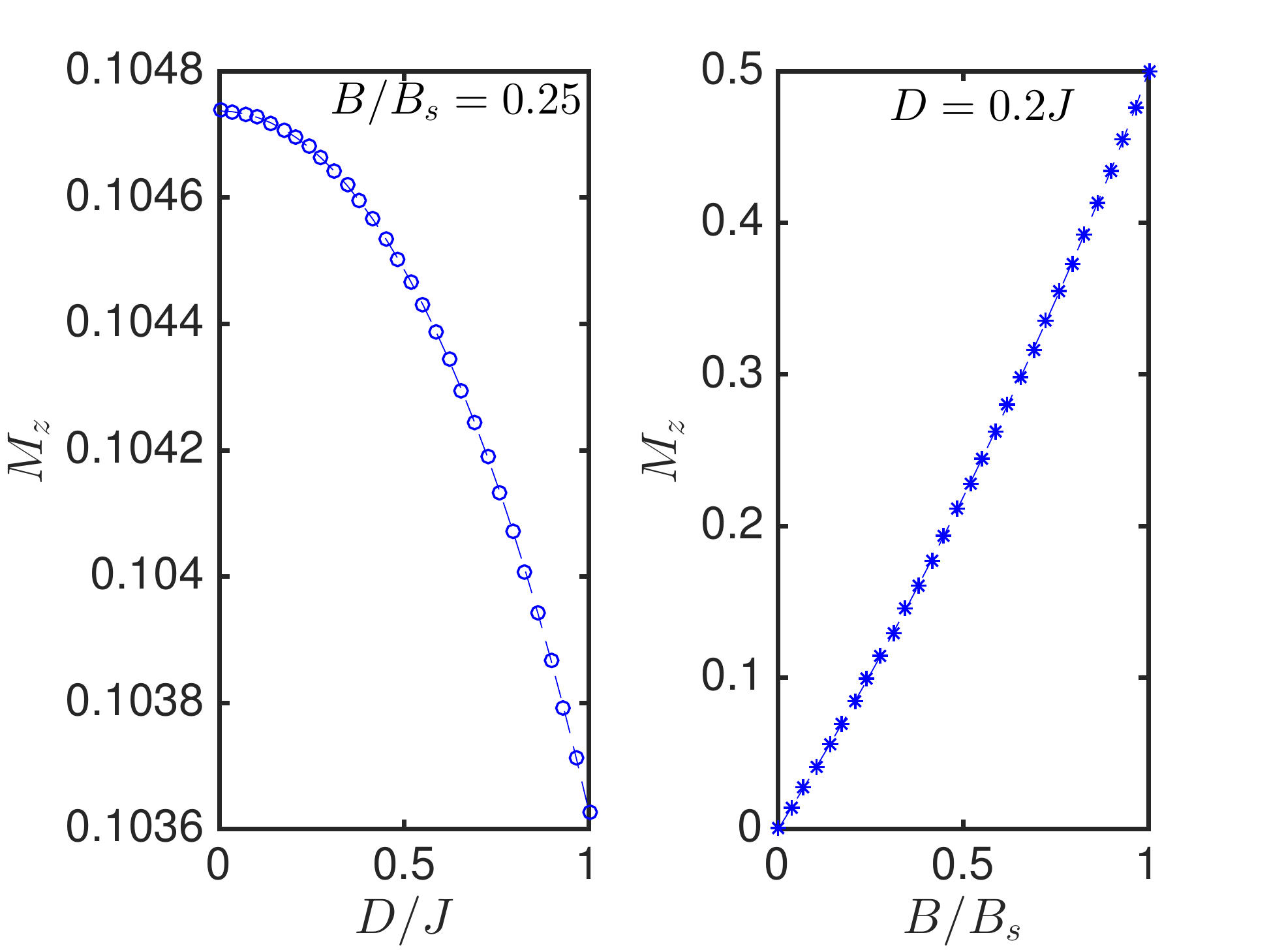}
\caption{Color online.  The  ground state average magnetization as function of  $D/J$ and $B/B_s$.}
\label{mag}
\end{figure}
 Figure \eqref{mag} shows the dependence of the ground state field-induced average magnetization as function of  the DMI and the magnetic field respectively. In the canted non-collinear antiferromagnetic phase an increase in DMI leads to a decrease in the magnetization. Although the DMI is an intrinsic property of  quantum magnetic materials, different materials have different DMI therefore the average magnetization will be different. The magnetic field has an opposite effect on the average magnetization. Starting from the collinear AFM  with zero  average magnetization the magnetic field increases the average magnetization to a maximum classical value $S=0.5$ at the saturation field, which is spontaneously formed in collinear FMs.
 \begin{figure}
\centering
\includegraphics[width=1\linewidth]{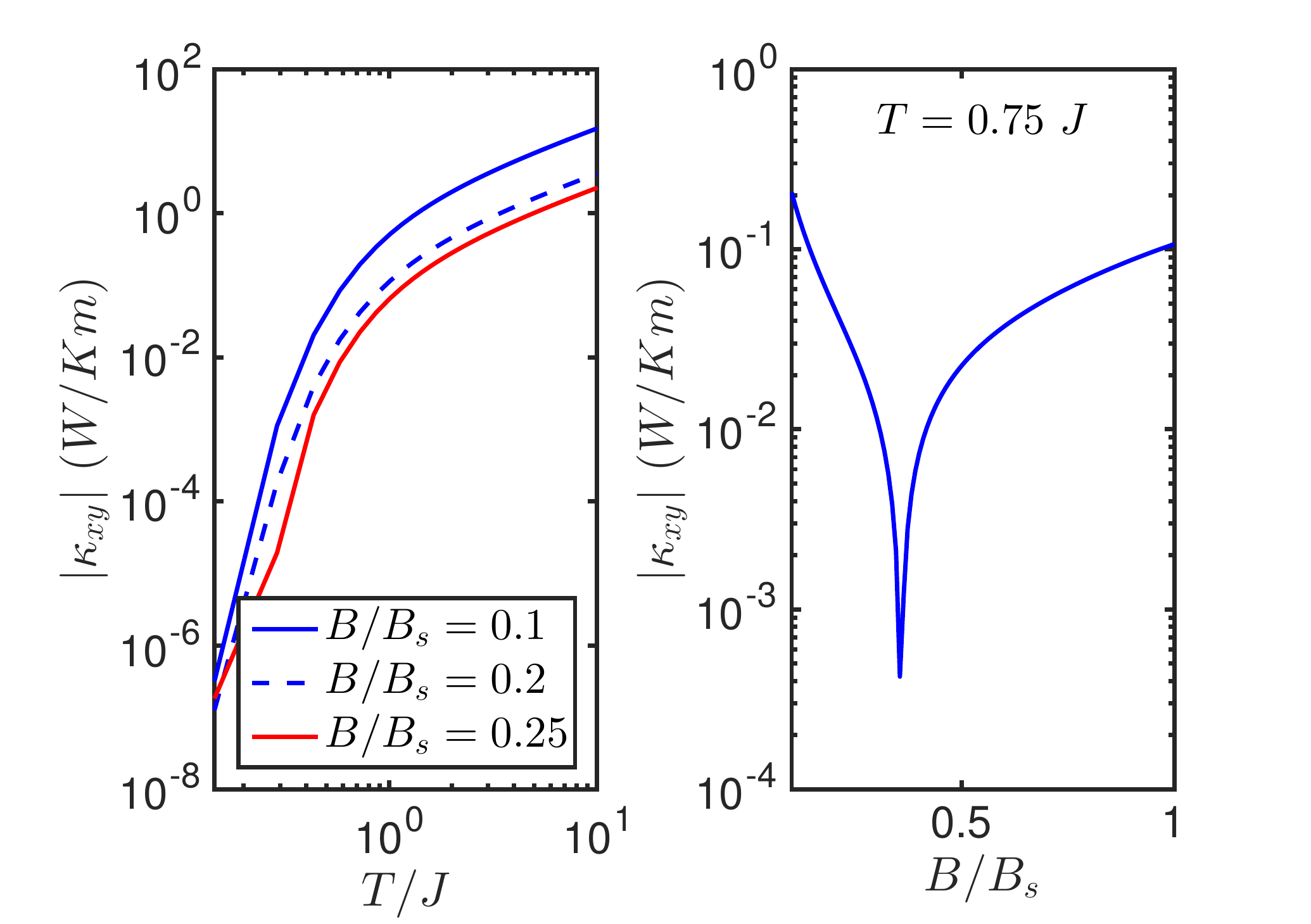}
\caption{Color online. (Left) Log-log plot of $|\kappa_{xy}|$ vs.  temperature  for various magnetic field in the canted AFM at  $D/J=0.2$.  (Right) Semilog plot of $|\kappa_{xy}|$ vs. magnetic field for $B\neq0$ at $T=0.75J$.  The right-hand side figure  shows a transition from canted AFM to collinear FM phase  at $B/B_s\sim 0.4$ (cf. Eq.~\ref{hl1}). }
\label{the1}
\end{figure}
\begin{figure}
\centering
\includegraphics[width=1\linewidth]{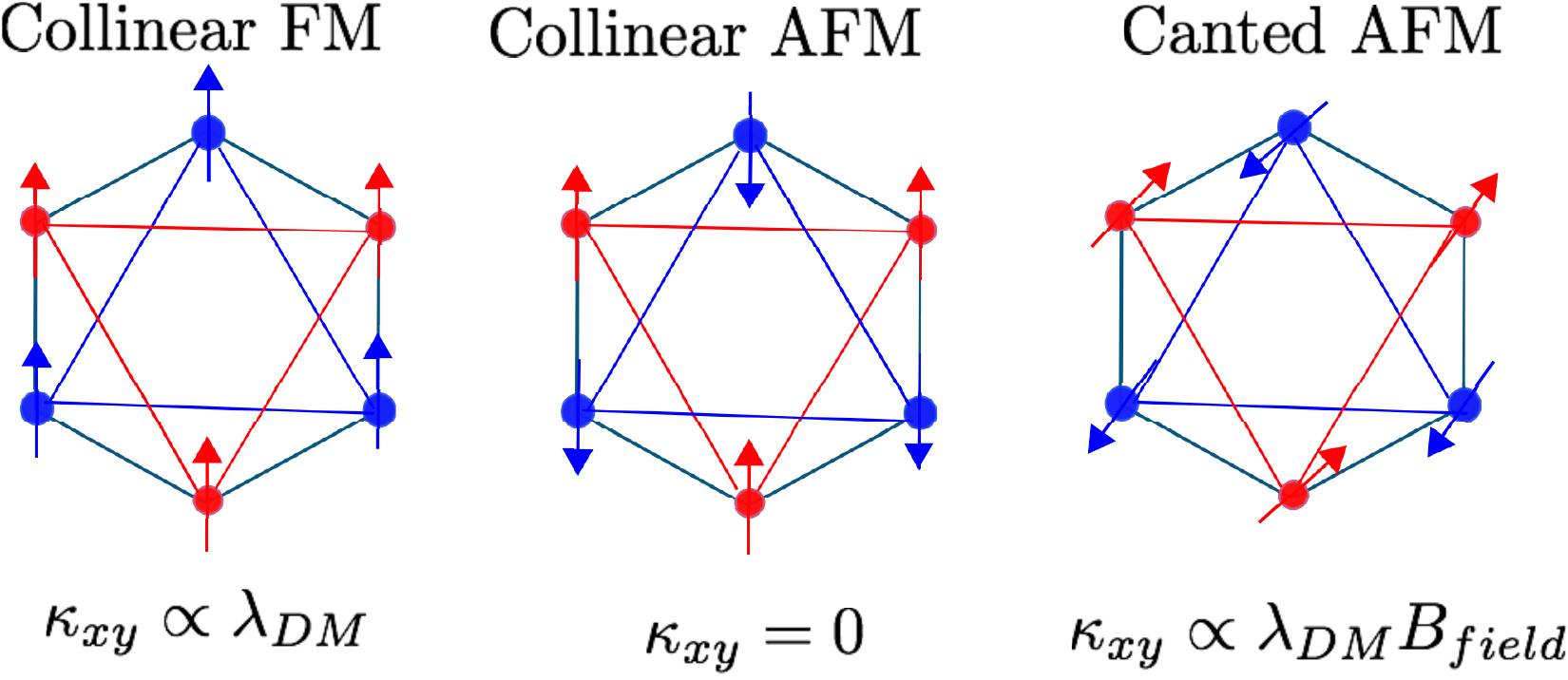}
 \caption{Color online. Schematics of thermal magnon Hall effect in bipartite honeycomb lattice. For collinear FM thermal Hall conductivity is induced by the DMI. In collinear AFM the thermal Hall conductivity vanishes, whereas in canted AFM the thermal Hall conductivity is induced by the magnetic field with nonzero DMI. }
\label{th_sum}
\end{figure}
 
  In real materials, a finite  Hall effect (thermal or quantum anomalous)  is usually assumed to be proportional to its magnetization, although no explicit relationship has been established. We therefore expect the present canted antiferromagnetic model to possess a finite thermal Hall effect. Plotted in Fig.~\ref{the1} is the thermal spin Hall conductivity as function of the temperature and the magnetic field. The temperature dependence of $\kappa_{xy}$ in the canted phase is not linear in the log-log plot, and thus does not show a power law unlike in ferromagnets. The field dependence of $\kappa_{xy}$ shows a phase transition from canted antiferromagnetic Haldane magnon insulator to collinear ferromagnetic Haldane magnon insulator.  Starting from $B\neq 0$  the collinear N\'eel order cant in the direction of the field, but at a certain critical field $B_c\sim 0.4B_s$ at $T=0.75J$ the conductivity function $\kappa_{xy}$ changes abruptly, which indicates a change of magnetic ordering as the system transits to  a collinear ferromagnetic insulator as shown in Fig.~\ref{the1} (right).  Such transition is expected because  at low temperatures only the states at $\bo={\bf \Gamma}$ have the maximum contribution to $\kappa_{xy}$ due to the Bose function \eqref{bose}. But the spin wave spectrum near $\bo={\bf \Gamma}$ is different for AFM and FM as shown in Eq. \ref{hl1}. Therefore the low-temperature dependence on $\kappa_{xy}$  should be different.   
The complete picture of thermal magnon Hall effect in bipartite honeycomb lattice is shown in Fig.~\ref{th_sum}. In collinear FMs, thermal magnon Hall effect is induced by the DMI. It should not require a magnetic field, but in some ferromagnetic materials the spin polarization is not along the direction of the out-of-plane DMI and a small Zeeman field might be needed to align the spins \cite{alex6,rc}.  Besides, a spontaneous magnetization is developed in FMs in the absence of magnetic field. In collinear AFMs the Hall response vanishes due to symmetry as well as the zero net magnetization.  A magnetic-field-induced canted non-collinear AFM  has finite Hall response  provided the DMI is present. We note that this result applies only to unfrustrated magnets. In frustrated magnets, thermal magnon Hall effect can be nonzero in the non-collinear regime even in the absence of the DMI \cite{sol2,sol3,sol4}.

\section{Conclusion}
 We have presented a complete investigation of topological magnon bands and thermal magnon Hall effect  in antiferromagnetic systems on the honeycomb lattice.  In contrast  to collinear ferromagnets  with spontaneous magnetization and DMI-induced Hall response, collinear antiferromagnets  have vanishing thermal Hall response due to time-reversal symmetry combined with lattice translation, which leads to doubly degenerate magnon dispersions with vanishing net magnetization. We  have shown that a finite thermal Hall response persists  in bipartite honeycomb antiferromagnetic compounds due to non-collinearity of the magnetic moments induced by an applied magnetic field.  The magnetic field not only induces a finite net magnetic moment, but  also breaks the degeneracy of the magnon dispersions and leads to a non-collinear antiferromagnetic Haldane magnon insulator. We note that non-collinear magnetic order can also occur in easy-axis antiferromagnets in a transverse in-plane magnetic field and easy-plane antiferromagnets in a longitudinal out-of-plane magnetic field.  These results have not been reported in the previous studies and therefore they provide an important experimental clue towards the realization of topological magnon properties in honeycomb antiferromagnetic materials. As mentioned in the Introduction there are several honeycomb antiferromagnetic materials which are potential candidates to confirm these results. The field-induced non-colllinear N\'eel order is also present in frustrated honeycomb antiferromagnets \cite{mak1, matt} --- an indication that the present results might be applicable to a wide range of honeycomb antiferromagnetic compounds. We note that the antiferromagnetic compounds MnF$_2$, FeF$_2$,  and CoO \cite{mac1,mac2,mac3} also have two-sublattice N\'eel order on a 3D cubic crystal structure. On any 2D surface (say $k_x$ and $k_y$) of these 3D magnetic structures  the magnon bands are similar to Eq.~\ref{hl} above.  Therefore, it is possible that the results of this paper could be verified in these 3D crystals as well if a DMI is present.

\acknowledgements
 Research at Perimeter Institute is supported by the Government of Canada through Industry Canada and by the Province of Ontario through the Ministry of Research
and Innovation.

\end{document}